\documentclass[12pt]{article}
\usepackage[english,german,french,polish]{babel}
\usepackage[T1]{fontenc}
\usepackage{amsfonts}

\begin{document}

\baselineskip 0.75cm
\topmargin -0.6in
\oddsidemargin -0.1in

\let\ni=\noindent

\renewcommand{\thefootnote}{\fnsymbol{footnote}}

\newcommand{\SM}{Standard Model }

\newcommand{\SMo}{Standard-Model }

\pagestyle {plain}

\setcounter{page}{1}

\pagestyle{empty}
~~~

\begin{flushright}
\end{flushright}

\vspace{0.3cm}

{\large\centerline{\bf Quarks and leptons which might be composite}}

\vspace{1.0cm}

{\centerline {\sc Wojciech Kr\'{o}likowski}}

\vspace{0.3cm}

{\centerline {\it Institute of Theoretical Physics, Faculty of Physics, University of Warsaw }}

{\centerline {\it Ho\.{z}a 69,~~PL--00--681 Warszawa, ~Poland}}

\vspace{3.0cm}

{\centerline{\bf Abstract}}

\vspace{0.3cm}

An option of composite quarks and leptons is briefly outlined, where elementary color-triplet quark-like fermions are bound with an elementary color-triplet isoscalar scalar boson due to the color coupling ${\bf 3}^*\!\times {\bf 3}^* \!\rightarrow {\bf 3}$ and ${\bf 3}^*\!\times {\bf 3} \!\rightarrow {\bf 1}$, respectively.

\vspace{0.6cm}

\ni PACS numbers: 12.15.Ff , 12.90.+b 

\vspace{0.8cm}

\ni May 2013

\vfill\eject

~~~
\pagestyle {plain}

\setcounter{page}{1}

\vspace{0.3cm}

Since the product of two color triplets can be reduced either to a color triplet or to a color singlet,

\begin{equation}
{\bf 3}^*\!\times {\bf 3}^* = {\bf 3} + {\bf6}^* \!\rightarrow {\bf 3} \;\;\;,\;\;\;{\bf 3}^*\!\times {\bf 3} = {\bf 1} + {\bf 8}^* \!\rightarrow {\bf 1}\;,
\end{equation} 

\ni one can imagine that the empirical quarks and leptons of three generations, $u_n = u, c, t\;,\; d_n = d, s, b$ and  $\nu_n = \nu_1, \nu_2, \nu_3\;,\; e_n = e^-, \mu^-, \tau^- \;(n = 1,2,3)$, are potentially bound states of the same elementary color-triplet constituents (preons) of two kinds: quark-like fermions $U_n$ and $D_n \,(n=1,2,3)$ of three generations with charges $Q_{U_n} = \frac{2}{3}$ and $Q_{D_n} = -\frac{1}{3}$ and an isoscalar scalar boson $S$ with charge $Q_S = -\frac{1}{3}$. In this case, the composite states are to be considered:

\begin{equation}
u_n = \left(D^c_n\,S^c \right) \;\,,\;\, d_n = \left(U^c_n\,S^c \right)  
\end{equation}

\ni and

\begin{equation}
\nu_n = \left(D^c_n\,S \right) \;\,,\,\; e_n = \left(U^c_n\,S \right) \,,
\end{equation}

\ni where $c$ denotes the charge conjugation.

Of course, the bound states (2) and (3) get the conventional charges

\begin{equation}
Q_{u_n} = \frac{1}{3} + \frac{1}{3} = \frac{2}{3} \;\,,\;\, Q_{d_n}  = -\frac{2}{3} + \frac{1}{3} = -\frac{1}{3}  \; 
\end{equation}

\ni and

\begin{equation}
Q_{\nu_n} = \frac{1}{3} - \frac{1}{3} = 0 \;\,,\,\;Q_{e_n} =  -\frac{2}{3} - \frac{1}{3} = -1\;, 
\end{equation}

\ni as well as the conventional baryon-minus-lepton numbers 

\begin{equation}
(B - L)_{u_n} = -\frac{1}{3} + \frac{2}{3} = \frac{1}{3} \;\,,\;\, (B - L)_{d_n}  = -\frac{1}{3} + \frac{2}{3} = \frac{1}{3}   
\end{equation}

\ni and

\begin{equation}
(B - L)_{\nu_n} = -\frac{1}{3} - \frac{2}{3} = -1 \;\,,\,\;(B - L)_{e_n} =  -\frac{1}{3} - \frac{2}{3} = -1  \;, 
\end{equation}

\ni respectively. The numbers (6) are to be identified with the baryon numbers $B_{u_n} = \frac{1}{3}$ and 
$B_{d_n} = \frac{1}{3}$ (then, $L_{u_n} = 0$ and $L_{d_n} = 0$), while the numbers (7) are interpreted as minus lepton numbers $L_{\nu_n} = 1$ and $L_{e_n} = 1$ (then $B_{\nu_n} = 0$ and $B_{e_n} = 0$). Here, $Q = I^3 + \frac{1}{2}(B-L)$, giving with the use of weak isospin $I^3_{U_n} = \frac{1}{2}$, $I^3_{D_n} = -\frac{1}{2}$, $I^3_{S} = 0$, the baryon-minus-lepton numbers $(B - L)_{U_n} = \frac{1}{3}$, $(B - L)_{D_n}  = \frac{1}{3}$, $(B - L)_{S} = -\frac{2}{3}$ applied in Eqs. (6) and (7).

An argument against  the elementary pointlike nature of quarks of the first generation, $u$ and $d$, can be the current mass $m_d$ significantly heavier than the current mass $m_u$ (by the factor 2.1 [1]). Since $u$ and $d$ do not differ by color, the color selfenergies are not expected to be responsible for such a mass difference. The electromagnetic selfenergies might play this role for elementary quarks of the first generation, if the charge-squared $Q^2_d$ were properly larger than $Q^2_u$ (but, in reality, $Q^2_d = \frac{1}{4}Q^2_u$ is smaller than $Q^2_u$). In contrast, for composite quarks (2) of the first generation, if their masses $m_u$ and $m_d$ are proportional to electromagnetic selfenergies  and so, propotional to

\begin{equation}
Q^2_{D_1} + Q^2_S = \left(\frac{1}{3}\right)^2 + \left(\frac{1}{3}\right)^2 = \frac{2}{9} \;\,,\;\, Q^2_{U_1} + Q^2_S =\left(\frac{2}{3}\right)^2 + \left(\frac{1}{3}\right)^2 = \frac{5}{9} ,
\end{equation} 

\ni  respectively, we infer that

\begin{equation} 
m_u {\bf :} m_d = \frac{2}{9}\, {\bf :}\, \frac{5}{9} = 1\, {\bf :}\,2.5 
\end{equation} 

\ni (here, electric radii of the involved preons are presumed to be reasonably equal). We can see that the resulting ratio $m_d/m_u = 2.5$ is not far away from the experimental ratio $m_d/m_u = 2.1$.

However,  the analogical argument applied to composite leptons (3) of the first generation is not consistent 
with the experimental smallness of $m_{\nu_1}/m_e$, because in this case there would be  
 
\begin{equation} 
m_{\nu_1} {\bf :}\, m_e = \frac{2}{9}\, {\bf :}\, \frac{5}{9} = 1\, {\bf :}\,2.5 
\end{equation} 

\ni like for quarks (2) of the first generation. Thus, a better understanding of the role of charges in close binding of charged preons is needed to solve the problem of $m_{\nu_1}/m_e \ll 1$ for our composite leptons. 

In order to fulfill the experimental requirement of $m_{\nu_1} \ll m_e$ and, at the same time, not to spoil our tentative explanation of the experimental ratio $m_d/m_u = 2.1$, we may try --- at a purely phenomenological level --- to replace in our argumentation the ansatz (8) by

\vspace{-0.4cm}

\begin{eqnarray}
\left(\frac{1}{3}\right)^2  +\left(\frac{1}{3}\right)^2 +\lambda \left(\frac{1}{3}\right)\left( \pm\frac{1}{3} \right)  & = &  \frac{2\pm\lambda}{9}\,\left\{\begin{array}{c}{{\rm  for\,quarks}\;u}\\ {{\rm  for\,leptons}\;\nu_1}
\end{array}\right. , \nonumber \\\, \left( \frac{2}{3} \right)^2\!\! +\left(\frac{1}{3}\right)^2\! +\lambda \left(\!\!-\frac{2}{3} \right)\left( \pm\frac{1}{3} \right) & = &  \!\frac{5\mp2\lambda}{9} \!\,\left\{\begin{array}{c}{{\rm  for\,quarks}\;d}\\ {{\rm  for\,leptons}\;e^-}\end{array}\right. \,,   
\end{eqnarray}

\vspace{0.2cm}

\ni respectively. Then, we have to put

\vspace{-0.2cm}

\begin{equation}
\lambda = 0\;{\rm to}\; 0.20\;\;{\rm for \; quarks }\;\; u,\,d
\end{equation}

\ni in order to obtain $m_d/m_u = 2.5\;{\rm\; to}\, 2.1$, or

\vspace{-0.3cm}

\begin{equation} 
\lambda \rightarrow 2 \;\;{\rm for\; leptons}\;\;\nu_1, e^-
\end{equation}

\ni to get $m_{\nu_1}/m_{e} \rightarrow 0$. The new term proportional to $\lambda$ on the lhs of Eqs. (11) corresponds to the phenomenological electromagnetic interaction between differently located charges 
$\frac{1}{3}$ and $\pm\frac{1}{3}$ within bound states $u$ and $\nu_1$, or between $-\frac{2}{3}$ and $\pm\frac{1}{3}$ within $d$ and $e^-$.

The plausible interpretation of ansatz (11) is that the role of interaction of differently located charges increases, when the inner radius of the bound states (involving these charges) decreases, leading to the coherent interaction of the sum of all involved charges with itself. This happens in the limit of $\lambda \rightarrow 2$ that is realized in the case of leptons. In this case, Eqs. (11) give

\vspace{-0.1cm}

\begin{eqnarray}
\left(\frac{1}{3}\right)^2\!\! +\left(\frac{1}{3}\right)^2\!\!+\lambda \left(\frac{1}{3}\right)\left( -\frac{1}{3} \right) \rightarrow \left(\frac{1}{3} - \frac{1}{3}\right)^{\!\!2} & = & 0 \;\;{{\rm  for\;leptons}\;\nu_1}, \nonumber \\\, 
\left( \frac{2}{3} \right)^2\!\! +\left(\frac{1}{3}\right)^2\!+\lambda \left(\!\!-\frac{2}{3} \right)\left(-\frac{1}{3}\right) \rightarrow \left(\frac{2}{3} + \frac{1}{3}\right)^{\!\!2} & = & 1 \;\;{{\rm  for\;leptons}\;\;e^-} .
\end{eqnarray}

\ni implying $m_{\nu_1}/m_e \rightarrow 0$. In the sense of this limit, leptons can be considered as the closest bound states of the charged preons involved. 

If the second-step composite states $(u_{\!1} u_{\!1} d_{\!1}\!),(u_{\!1} d_{\!1} d_{\!1}\!)$ and $(u_{\!1} d_{\!1}^c),\frac{1}{\sqrt{2}}[(u_{\!1} u_{\!1}^c)-(d_{\!1} d_{\!1}^c)], (u_{\!1}^c d_{\!1})$ are empirical nucleons and pions, respectively, then the bound states such as the first-step "primordial nucleons"\,$(U_1 U_1 D_1) , (U_1 D_1 D_1)$ and "primordial pions"\,$(U_1 D_1^c),\frac{1}{\sqrt{2}}[(U_1 U_1^c)-(D_1 D_1^c)]$, $(U_1^c D_1)$ --- when they exist --- contribute an {\it embarras de richesse} to our model, even to its first generation $n=1$. A radical way out of such a difficulty might be the nonexistence of these bound states as physical ones, if they were conjectured to be too strongly attracted (too closely bound) by color binary interactions of the quark-like fermions $ U_1 , D_1, U^c_1 , D^c_1$. In this case, however, the constructive attraction of color-triplet isoscalar scalar $S$ with $U_n , D_n$ and $U_n^c , D_n^c$ would be necessary to compose the empirical quarks $u_n = (D_n^c S^c)$ , $d_n = (U_n^c S^c)$ and leptons $\nu_n = (U_n^c S)$ , $e_n = (D_n^c S)\;(n=1,2,3)$, the former having to be color constituents of observed hadrons.

For a proposal of two-body Dirac and Klein-Gordon equation {\it cf.} Ref. [2]; the model potential therein might be $V = -(1\;{\rm or}\;2)\alpha_s/r + (\beta_s\;{\rm or}\;\gamma_s)r$ with $\beta_s$ or $\gamma_s$ depending on the generation $n$ of $U_n , D_n$ or $U_n^c , D_n^c$, respectively, interacting with $S$. It is interesting to notice that for $V = -2\alpha_s/r$ with $\alpha_s = 1$ and equal masses this two-body equation leads to a massless ground state in the generation $n=1$. For $V = -2\alpha_s/r$ with $\alpha_s > 1$ the energy spectrum becomes complex, so it is not longer physical, while with $\alpha_s \leq 1$ it remains real. Analogical effects appear for two-body double-Dirac equation.

Now, in the framework of our composite model of quarks and leptons, we will concentrate on their quark-like fermion constituents $U_n$ and $D_n$ in three generations ($n = 1,2,3$), in order to explain the phenomenon of three generations of empirical quarks and leptons.

To this end, it is natural to presume that these fermions should carry an odd number of Dirac bispinor indices $\alpha_1, \alpha_2, \ldots, \alpha_N (N = 2n-1 = 1,3,5,...\,,\; n=1,2,3,...)$ and the (suppressed) color-triplet label:


\begin{equation}
U_n(x) \leftrightarrow (U^{(n)}_{\alpha_1, \alpha_2, \ldots, \alpha_N}(x) \;\; , \;\; D_n(x) \leftrightarrow (D^{(n)}_{\alpha_1, \alpha_2, \ldots, \alpha_N}(x)\,,
\end{equation}

\ni Here, one of Dirac bispinor indices $\alpha_i\; (i = 1,2,\ldots,N)$, in fact $\alpha_1$, is correlated with the (suppressed) color-triplet label and so, is distinguished from the remaining $\alpha_2,\alpha_3,\ldots,\alpha_ N$ which can be considered as undistinguishable from each other. Thus, the Dirac bispinor indices $\alpha_2, \alpha_3, \ldots, \alpha_N$ behave as physical objects ("intrinsic partons") obeying Fermi statistics along with Pauli exclusion principle ("intrinsic Pauli principle") requiring them to be fully antisymmetrized. This implies that N can be equal to 1,3,5 only (and so, $n$ equal to 1,2,3 only), since any $\alpha_i$ assumes four values 1,2,3,4. In addition, the total spin of $U_n$ and $D_n$ is reduced to spin $\frac{1}{2}$ correlated with the distinguished bispinor index $\alpha_1$. Hence, we can conclude that in Nature there are exactly three generations of quark-like fermionic preons $U_n$ and $D_n$, and, in consequence, there are three generations of composite quarks and leptons ({\it cf.} Ref. [3] for the option of three generations of elementary quarks and leptons).

Explicitly, the correspondence (15) reads

\begin{eqnarray}
U_{1 \alpha_1}(x) & = & U^{(1)}_{\alpha_1}(x) , \nonumber \\
U_{2 \alpha_1}(x) & = & \frac{1}{4}\left(C^{-1} \gamma^5 \right)_{\alpha_2 \alpha_3}U^{(2)}_{\alpha_1 \alpha_2 \alpha_3}(x) = U^{(2)}_{\alpha_1 1 2}(x) = U^{(2)}_{\alpha_1 3 4}(x) , \nonumber \\
U_{3 \alpha_1}(x) & = & \frac{1}{24}\varepsilon_{ \alpha_ 2 \alpha_3 \alpha_4 \alpha_5}U^{(3)}_{\alpha_1 \alpha_2 \alpha_3 \alpha_4 \alpha_5}(x) = U^{(3)}_{\alpha_1 1 2 3 4}(x) , \nonumber \\
\end{eqnarray}

\ni and, analogically, for $D_{n \alpha_1}(x)\;(n=1,2,3)$. In addition, $U^{(2)}_{ \alpha_1 1 3}(x) = U^{(2)}_{\alpha_1 2 4}(x)$, $U^{(2)}_{ \alpha_1 1 4}(x) = U^{(2)}_{\alpha_1 2 3}(x)$ and, analogically, for  $D^{(2)}_{\alpha_1 \alpha_2 \alpha_3}(x)\;(\alpha_2 \alpha_3 = 13,24,14\;{\rm and}\; 23)$. Here, the probability interpretation and relativity of quantum theory is applied.


\vspace{0.5cm}

{\centerline{\bf References}}

\vspace{0.2cm}

\baselineskip 0.75cm


\begin{description}

\item{[1]}~J.Beringer~{\it et al.} (Particle Data Group), {\it Phys. Rev.} {\bf D 86}, 010001 (2012).

\item{[2]}~W. Kr\'{o}likowski, {\it Acta Phys. Polon.} {\bf B 10}, 735 (1979). 

\item{[3]}~W.~Kr\'{o}likowski, {\it Acta Phys. Polon.} {\bf B 23}, 933 (1992); {\it Phys. Rev.}, {\bf D 45}, 3222 (1992); {\it Acta Phys. Polon.}, {\bf B 33}, 2559 (2002).

\vspace{0.2cm}

\end{description}

\vfill\eject

\end{document}